\documentclass[twocolumn,twocolappendix,tighten]{aastex701}

\shorttitle{Multiple protostellar outflows}
\shortauthors{Hirano et al.}

\received{February 24, 2026}
\revised{April 13, 2026}
\accepted{April 20, 2026}

\submitjournal{ApJ}

\graphicspath{{./}{figures/}}

\newcommand{\msun}{M_\odot}

\newcommand{\msunyr}{M_\odot\,{\rm yr}^{-1}}
\newcommand{\K}{\rm K}
\newcommand{\cc}{{\rm cm^{-3}}}
\newcommand{\kms}{{\rm km\,s^{-1}}}

\newcommand{\tps}{t_{\rm ps}}
\newcommand{\Mps}{M_{\rm ps}}
\newcommand{\Mdisk}{M_{\rm disk}}
\newcommand{\Rdisk}{R_{\rm disk}}

\newcommand{\zdw}{z_{\rm DW}}
\newcommand{\rsf}{r_{\rm SF}}

\begin{document}


\title{Multiple protostellar outflows from a single protostar with a misaligned disk}

\correspondingauthor{Shingo Hirano}

\author[orcid=0000-0002-4317-767X,gname='Shingo',sname='Hirano']{Shingo Hirano}
\affiliation{Department of Applied Physics, Faculty of Engineering, Kanagawa University, Kanagawa 221-0802, Japan}
\affiliation{Department of Astronomy, School of Science, University of Tokyo, Tokyo 113-0033, Japan}
\email[show]{shingo-hirano@kanagawa-u.ac.jp}

\author[orcid=0000-0003-3283-6884,gname='Yuri',sname='Aikawa']{Yuri Aikawa}
\affiliation{Department of Astronomy, School of Science, University of Tokyo, Tokyo 113-0033, Japan}
\email{aikawa@astron.s.u-tokyo.ac.jp}

\author[orcid=0000-0002-0963-0872,gname='Masahiro',sname='Machida']{Masahiro N. Machida}
\affiliation{Department of Earth and Planetary Sciences, Faculty of Science, Kyushu University, Fukuoka 819-0395, Japan}
\email{machida.masahiro.018@m.kyushu-u.ac.jp}


\begin{abstract}
We investigate how misalignment between the core angular momentum and the large-scale magnetic field affects protostellar outflows, and whether a single protostellar system can drive multiple outflow components.
We perform three-dimensional nonideal magnetohydrodynamic simulations of magnetized rotating cores, focusing on the formation of a protostar, a circumstellar disk, and magnetically driven outflows.
The initial angle between the core angular-momentum vector and the magnetic field is systematically varied from $0^\circ$ to $90^\circ$.
All models launch a classical magnetocentrifugal disk wind (DW) roughly along the local disk normal.
For large misalignment, the system also develops a spiralflow (SF) component that propagates parallel to the disk plane.
In a representative model with a $60^\circ$ misalignment, the outflow transitions from a DW-dominated to an SF-dominated phase, with the SF becoming more massive and more extended than the DW, and the two components intermittently coexisting.
Across the model suite, the maximum mass and size ratios of SF to DW, as well as the relative lifetimes of the two components, increase for misalignment angles $\gtrsim60^\circ$.
We propose that secondary, misaligned outflows (or their fossil remnants) observed in some protostellar systems can be interpreted as the SF component, while the main bipolar outflow traces the DW from the same misaligned system.
\end{abstract}

\keywords{\uat{Magnetohydrodynamical simulations}{1966} --- \uat{Star formation}{1621} --- \uat{Protostars}{1302} --- \uat{Circumstellar disks}{235} --- \uat{Stellar jets}{1607} --- \uat{Circumstellar envelopes}{237}}


\section{Introduction} \label{sec:intro}

Protostellar outflows are a key mechanism of feedback during the earliest stages of low-mass star formation.
They remove mass and angular momentum from the circumstellar disk and infalling envelope, help to regulate the final stellar mass, and strongly affect the surrounding environment.

High-angular-resolution observations with Atacama Large Millimeter/submillimeter Array (ALMA) have recently revealed complex, multi-component outflows from young protostars.
In several single protostellar systems, in addition to the main bipolar outflow, a secondary outflow-like structure is detected at a different position angle and with distinct kinematics, often referred to as a secondary flow or misaligned outflow (or its fossil) \citep[e.g.,][]{Okoda2021,Sato2023,Sai2024}. 
In many of these sources, the projected axes of the primary and secondary components are nearly orthogonal, with the secondary flow emerging closer to the disk plane or at a large offset angle from the main bipolar outflow.
Such a configuration suggests that the two components are physically related yet follow different launch directions around the same central source.

In the classical theoretical picture, outflows are launched by magnetocentrifugal force along open magnetic field lines anchored in a protostellar disk \citep{Tsukamoto2023PPVII}.
Such magnetocentrifugal disk winds (DW) naturally produce collimated, bipolar outflows roughly aligned with the disk rotation axis \citep[e.g.,][]{BlandfordPayne1982, PudritzNorman1986, Shu1994}.
Modern three-dimensional magnetohydrodynamic (MHD) simulations have demonstrated that DW arise robustly in collapsing, rotating, magnetized cores and have clarified how magnetic braking, nonideal MHD effects, and thermodynamics influence the outflow efficiency and morphology \citep[e.g.,][]{Tomida2015,Tsukamoto2015}.
In many of these previous studies, however, the initial configuration was chosen such that the rotation axis and the large-scale magnetic field were aligned, or only weakly misaligned.

In a realistic star formation process, however, such alignment is not guaranteed.
Previous numerical studies have shown that misalignment between the rotation axis and the magnetic field can weaken magnetic braking, promote the formation of sizable circumstellar disks, and lead to warped or precessing disks \citep[e.g.,][]{HennebelleCiardi2009,Joos2012,Tsukamoto2018}.

Nevertheless, the impact of such misalignment on the multiplicity and geometry of outflows remains less well explored.
In particular, it is not yet clear under what conditions a single protostellar system can produce more than one pair of bipolar outflow lobes with clearly different axes and position angles.

Misalignment enables magnetic field lines to penetrate the disk from the side rather than along the disk normal, producing spiral magnetic structures that lie nearly parallel to the disk plane \citep[see Figure~8 of][]{Machida2006}.
Along these side-penetrating field lines, the disk rotation winds up the field within the disk plane, generating a toroidal component. 
A distinct spiralflow (SF) is launched from the disk and accelerated along the twisted field lines by magnetic and centrifugal effects.
In this sense, the resulting flow geometry is qualitatively analogous to the Parker spiral structure of the solar wind \citep{Parker1958}, although the SF is powered by magnetic energy stored in the twisted field rather than by the solar-wind acceleration mechanism \citep[e.g.,][]{Weber1967,Spruit1996,Spruit2009}.

Spiral-shaped, magnetically driven outflows have already been identified in numerical simulations of misaligned collapsing cores.
\citet{MatsumotoHanawa2011} and \citet{Matsumoto2017} showed that twisted magnetic fields can launch an SF in addition to a more classical magnetocentrifugal wind, and \citet{Machida2006} investigated magnetic-field morphologies for inclined rotators.  
However, these studies did not systematically quantify how SFs coexist with or compete with a magnetocentrifugal DW throughout protostellar evolution, nor did they directly connect these SFs to the multiple outflows now being revealed by ALMA.

In this work, we address whether a single protostellar system can drive multiple outflow components using a suite of three-dimensional nonideal MHD simulations of collapsing, rotating dense cores with various misalignment angles between the core angular-momentum vector and the large-scale magnetic field.
Our numerical setup follows \citet{Hirano2025factor3}, and we systematically vary the initial misalignment angle $\theta_0$ from $0^\circ$ (aligned) to $90^\circ$ (orthogonal).
We show that, in addition to the classical magnetocentrifugal DW, the misaligned models develop a distinct SF component launched roughly along the disk plane, nearly perpendicular to the propagation direction of the primary bipolar wind.
Crucially, we find that DW and SF are not mutually exclusive: both can be driven by the same protostellar disk within a single collapsing cloud core, with their relative importance determined by the degree of misalignment.

By quantifying how the relative strength and lifetime of the SF depend on the misalignment angle $\theta_0$, we provide a concrete theoretical model in which multiple outflow components naturally arise from a single misaligned protostellar system.  
We propose that the observed secondary outflows can be interpreted as the projected manifestation of the SF component, while the main bipolar outflow traces the DW.  
Our results thus link the SFs previously identified in simulations \citep[e.g.,][]{MatsumotoHanawa2011,Matsumoto2017,Machida2020} with the recently observed multiple outflows, and highlight misaligned magnetic fields as a promising origin of such ``multiflow'' outflows with multiple components.

\section{Methods} \label{sec:methods}

\subsection{Simulation}
\label{sec:method_sim}

We perform three-dimensional nonideal magnetohydrodynamic (MHD) simulations of collapsing, rotating, magnetized dense cores that form a protostar, a circumstellar disk, and associated outflows.  
The numerical scheme, microphysics, and initial conditions follow \citet{Hirano2025factor3}, and we briefly summarize the setup relevant to this work.

The simulations solve the resistive MHD equations with self-gravity on a nested-grid mesh.
Ohmic dissipation is included using the same prescription as in \citet{Hirano2025factor3}, following the implementation of \citet{Machida2007ApJ}; specifically, the induction equation is solved with the Ohmic term in the form $\eta_{\rm O} \nabla^2 \mathbf{B}$. Appendix~\ref{sec:app:ohmic} addresses the implication of this approximation.
The gas thermodynamics are treated with a barotropic equation of state.
The finest spatial resolution reaches $\Delta x_{\rm min} = 0.38$\,au, which is sufficient to resolve the inner disk and the outflow launching region around the protostar.

The initial condition is a dense core similar to a Bonnor--Ebert sphere with mass $M_{\rm core} = 3.2\,\msun$ and central number density $n_{\rm c,0} = 5\times10^5\,\cc$, embedded in a lower-density ambient medium.  
The core is in solid-body rotation with angular velocity $\Omega_0 = 1.46 \times 10^{-13}\,{\rm s}^{-1}$ and is threaded by a uniform magnetic field $B_0 = 4.9 \times 10^{-5}$\,G.
Under these conditions, the ratios of thermal and rotational energies to the gravitational energy are $\alpha_0 = 0.4$ and $\beta_0=0.02$.
The mass-to-flux ratio normalized by the critical value $(4 \pi^2 G)^{-1/2}$ is $\mu_0 = 3$.
The angle between $\Omega_0$ and $B_0$ defines the misalignment parameter $\theta_0$.  
In this work, we consider a series of magnetized models with $\theta_0 = \{ 0^\circ,\ 15^\circ,\ 30^\circ,\ 45^\circ,\ 60^\circ,\ 75^\circ,\ 90^\circ \}$, which we refer to as T00--T90, respectively.

Protostar formation is treated using a sink particle that is introduced once the gas number density exceeds a threshold $n_{\rm sink} = 10^{13}\,\cc$.
We define the protostellar time as $\tps = 0$ at sink creation and follow the subsequent evolution up to $\tps = 10^5\ {\rm yr}$.
Snapshot data are stored every $10^3\ {\rm yr}$.
At each snapshot, we record the protostellar mass $\Mps$ (i.e., the sink-particle mass) and measure the disk mass $\Mdisk$ by integrating the gas in a dense, rotationally supported region around the sink.

\subsection{Outflow analysis}
\label{sec:method_analysis}

For the outflow analysis, we first identify the circumstellar disk and define a characteristic disk radius.
The disk is taken to be rotationally supported, dense material within a compact region around the protostar, selected by the criteria $v_{\rm rot} \ge 0.95\,v_{\rm Kep}$, $v_{\rm rot} \ge 10\,|v_{\rm rad}|$, $n \ge 10^{10}\,\cc$, and $r \le 150\,{\rm au}$, where $v_{\rm rot}$ and $v_{\rm rad}$ are the azimuthal and radial velocity components, $v_{\rm Kep}$ is the local Keplerian velocity, $n$ is the gas number density, and $r$ is the spherical radius from the protostar.
We then define the disk radius $\Rdisk$ as the maximum distance from the sink among the gas cells that satisfy these criteria.

At each snapshot, we construct a local ``disk-frame'' coordinate system $(x', y', z')$ in which the $z'$-axis is aligned with the instantaneous angular-momentum vector of the dense disk.
We first translate positions to sink-centered coordinates, $\mathbf{r}'=\mathbf{r}-\mathbf{r}_{\rm sink}$, and use velocities relative to the sink, $\mathbf{v}'=\mathbf{v}-\mathbf{v}_{\rm sink}$.
We then compute the specific angular momentum of gas within $r (=\sqrt{x'^2+y'^2}) < \Rdisk$ and above a density threshold $n > n_{\rm disk} = 10^{10}\,\cc$, and define its unit vector as $\hat{\mathbf{J}}_{\rm disk}$.
We apply a rotation matrix that maps $\hat{\mathbf{J}}_{\rm disk}$ onto the $z'$-axis.
In the resulting disk frame, the disk lies approximately in the $x'$--$y'$ plane and the local disk axis is aligned with $z'$.

In the disk frame, we classify gas into three kinematic categories: disk, inflow from the envelope to the disk, and outflow from the disk.
The disk mask is defined by applying the same kinematic and density criteria as above within $r \le \Rdisk$,
while gas that does not satisfy the disk criteria is further separated into inflowing and outflowing components based on the sign and magnitude of $v_{\rm rad}$ as described below.

Gas participating in outflows is selected by requiring radial motion away from the protostar and relatively low density.
We define the outflow mask by $v_{\rm rad} > 1.0\ \kms$ and $10^3 < n\,\cc < 10^{8}$,\footnote{The lower density limit is the critical density of CO, $\sim\!10^{3}\,\cc$.} and only cells satisfying this condition and lying outside the dense disk are treated as outflow material.
All remaining gas is regarded as an infalling envelope or quasi-static core material and is not analyzed further in this study.

\begin{figure*}[t!]
\includegraphics[width=1.0\linewidth]{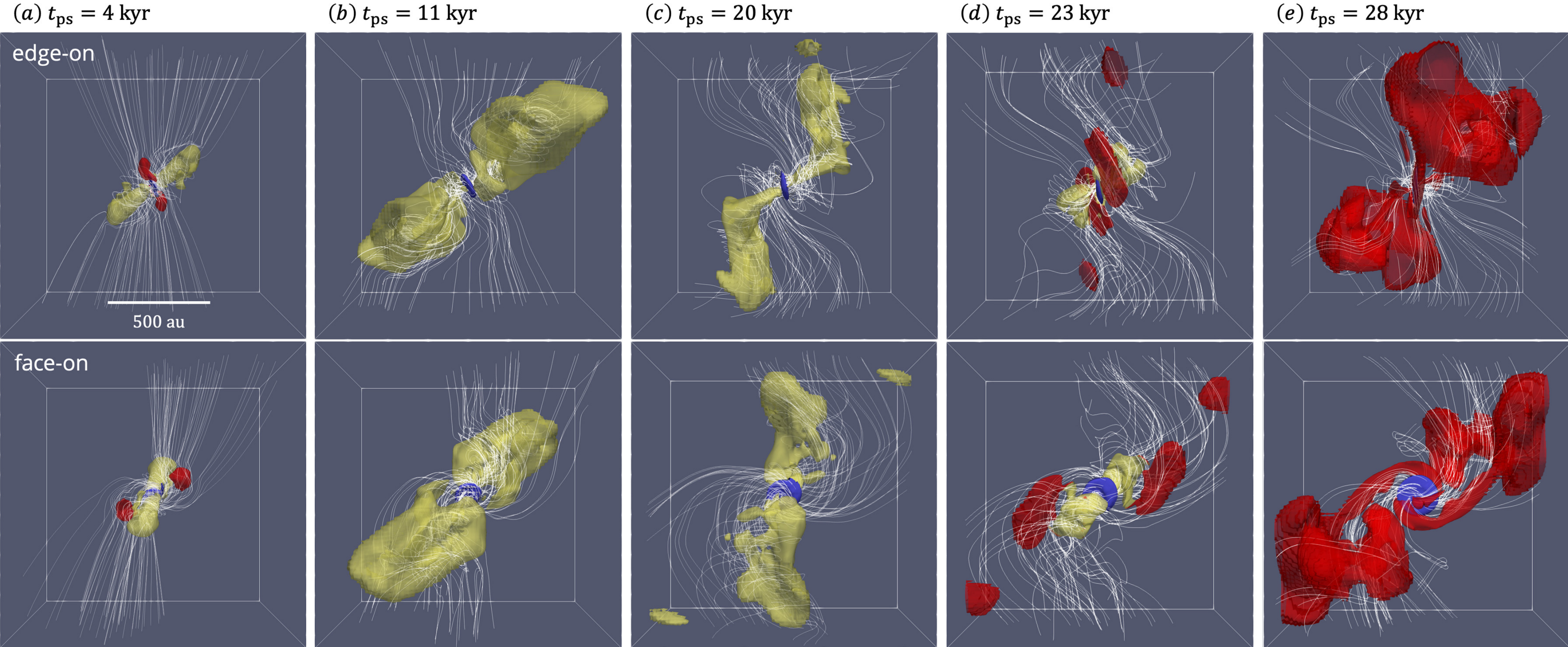}
\caption{
Three-dimensional structure of the magnetized protostellar system in the misaligned model T60 with an initial angle of $\theta_0 = 60^\circ$.
We show the density contours of the disk (blue), the disk wind (DW; yellow), and the spiralflow component (SF; red).
The white lines show the magnetic field lines.
}
\label{fig:3dview_10panels}
\end{figure*}

\begin{figure*}[t!]
  \begin{minipage}[t]{0.49\linewidth}
    \centering
    \includegraphics[width=\linewidth]{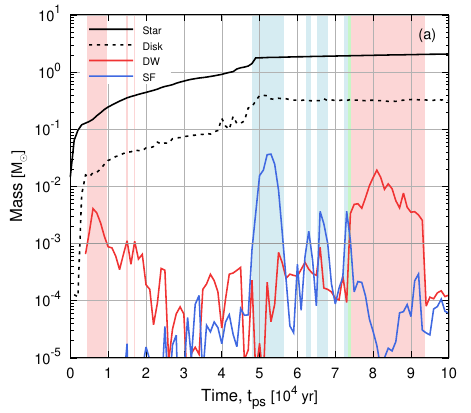}
  \end{minipage}
  \hfill
  \begin{minipage}[t]{0.49\linewidth}
    \centering
    \includegraphics[width=\linewidth]{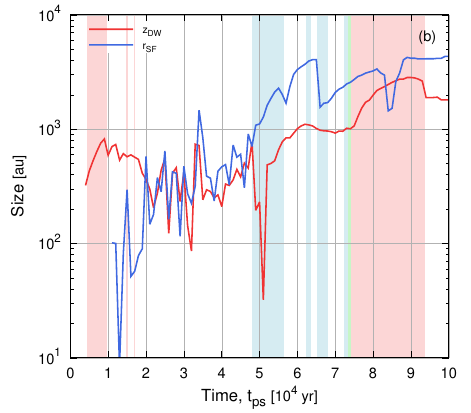}
  \end{minipage}
\caption{
Time evolution of the outflow properties in the misaligned model T60.
Panel (a): Masses of the protostar, circumstellar disk, disk wind (DW), and spiralflow (SF) components as functions of time after protostar formation, $\tps$.
Panel (b): Corresponding characteristic extents of the DW and SF, where $\zdw$ is the maximum vertical distance along the disk axis and $\rsf$ is the maximum radial distance in the disk plane.
In both panels, the colored background indicates phases during which each outflow component is strongly present: blue, red, and green shading correspond to periods when the DW mass, the SF mass, and both components simultaneously exceed a fiducial threshold of $10^{-3}\,\msun$, respectively, using the same criterion as in Figure~\ref{fig:theta-values}(b).  
}
\label{fig:tps-values}
\end{figure*}

Within the outflow, we separate the DW and SF components using a simple geometric classifier based on the polar angle from the disk rotation axis,
\begin{equation}
\theta_{\rm cri} = \arctan \left( \frac{\sqrt{x'^2 + y'^2}}{|z'|} \right) = \arctan (2) \simeq 63.4^\circ ,
\end{equation}
where $\theta$ is measured from the disk normal ($\theta=0$ and $\pi$ at the two poles).
We define the DW region as a bipolar conical volume satisfying $0 < \theta < \theta_{\rm cri}$ or $\pi - \theta_{\rm cri} < \theta < \pi$, while the SF region is outflowing gas located outside this double cone.
This criterion is adopted for analysis convenience and does not, by itself, imply a unique launching condition or physical mechanism.
Values of $\theta_{\rm cri}$ much smaller than $\arctan(2)$ tend to classify laterally broadened DW material as SF, whereas much larger values miss a substantial fraction of the SF component.
After exploring several choices, we adopt $\theta_{\rm cri} = \arctan(2)$ as a representative value that cleanly separates the two components in all models.

With this choice, DW cells correspond to a relatively collimated, axis-oriented flow, whereas SF cells represent a broader, disk-plane–oriented flow.
The same thresholds are consistently applied to all snapshots and misalignment models.

\begin{figure*}[t!]
  \begin{minipage}[t]{0.49\linewidth}
    \centering
    \includegraphics[width=\linewidth]{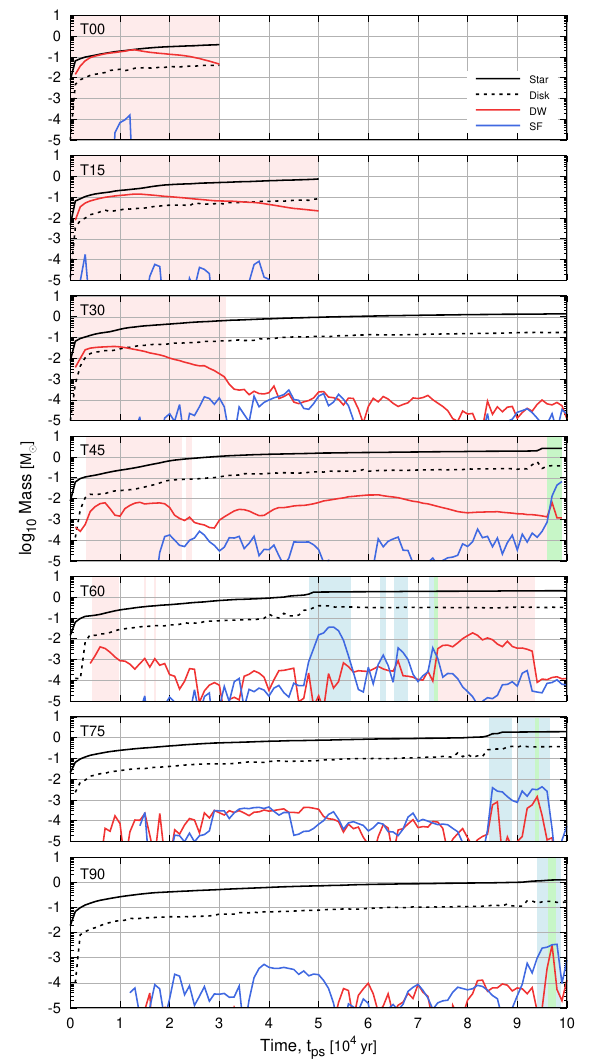}
  \end{minipage}
  \hfill
  \begin{minipage}[t]{0.49\linewidth}
    \centering
    \includegraphics[width=\linewidth]{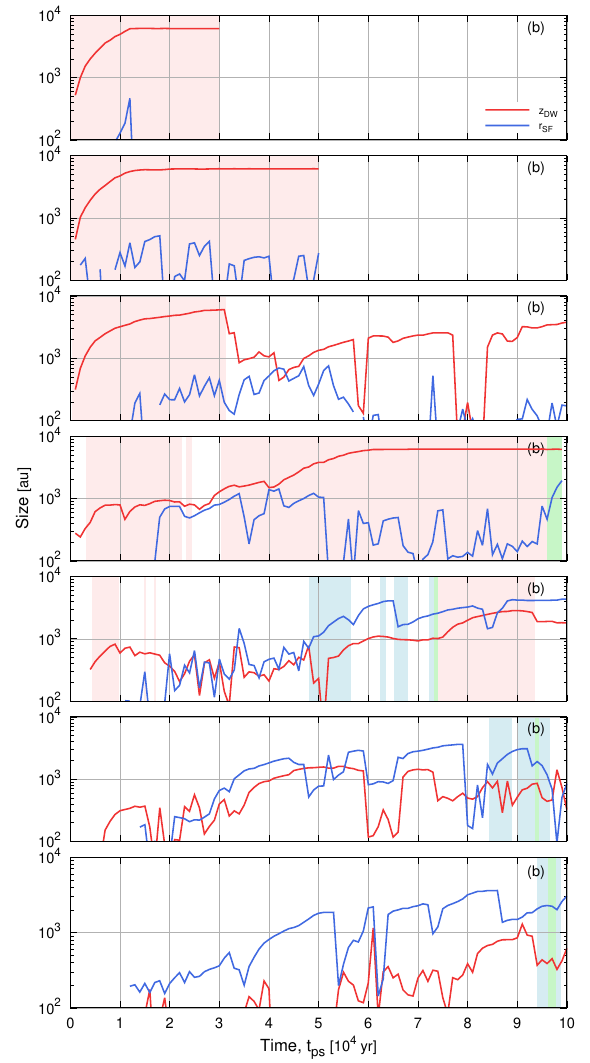}
  \end{minipage}
\caption{
Time evolution of the outflow properties for all misalignment models T00--T90.
This figure generalizes the single-model evolution shown in Figure~\ref{fig:tps-values} to the full range of misalignment angles and provides the time-series data from which the summary diagnostics in Figure~\ref{fig:theta-values} are derived.
}
\label{fig:tps-values_T00-90}
\end{figure*}

\begin{figure}[t!]
\centering
\includegraphics[width=\linewidth]{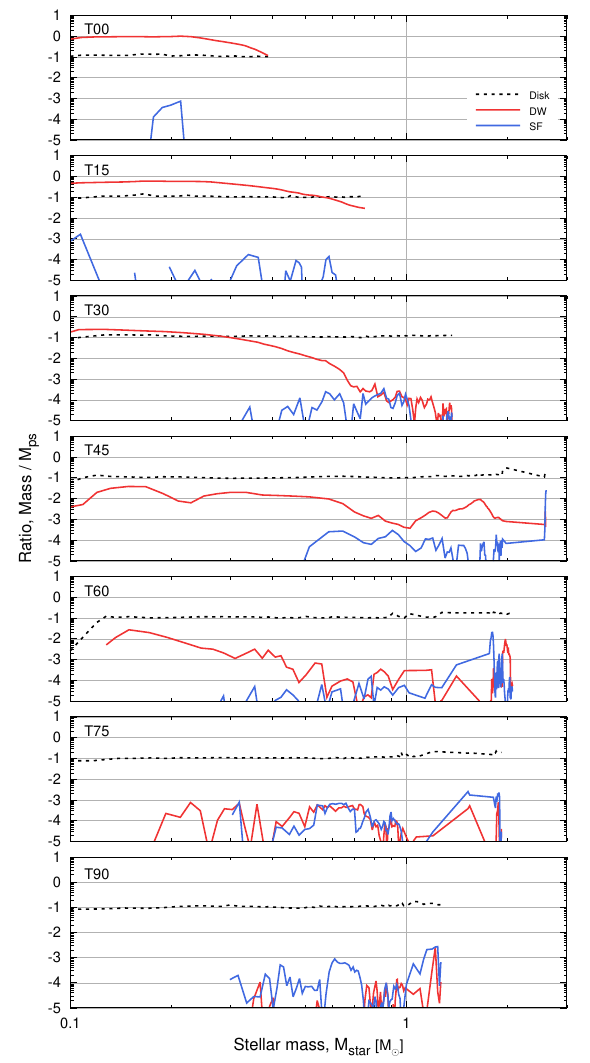}
\caption{
Mass fractions of the circumstellar components (disk, DW, and SF) as functions of the protostellar mass for models T00--T90 (from top to bottom).
}
\label{fig:mps-mass_T00-90}
\end{figure}

\section{Results} \label{sec:res}

We first describe the multiflow structure in the representative misaligned model with $\theta_0 = 60^\circ$ (model T60).
We then quantify how the relative importance of the DW and SF components evolves in time and how it depends on the initial misalignment angle.

\subsection{Multiflow outflows in the misaligned model T60}
\label{sec:res:T60}

Figure~\ref{fig:3dview_10panels} presents the main components (disk, DW, and SF) and magnetic-field structure around the protostar in model T60.
We highlight the two distinct outflow components identified by our kinematic classification: the DW, which is launched roughly along the disk axis, and the SF, which propagates preferentially along the disk plane.
Both components clearly originate from the same disk, but follow different spatial directions set by the misaligned and twisted magnetic field.

Figure~\ref{fig:tps-values} shows the temporal evolution of the mass budget in model T60.
In panel~(a) we plot the protostellar mass $\Mps$, the disk mass $\Mdisk$, and the masses in the DW and SF components, $M_{\rm DW}$ and $M_{\rm SF}$, as functions of the protostellar time $\tps$.
We highlight these phases by shading the background whenever the DW mass, the SF mass, or both exceed $10^{-3}\,\msun$ (blue, red, and green shading, respectively).
Shortly after $\tps=0$, the DW launches, and its mass rapidly increases, while the SF component remains negligible.
This early phase resembles the classical DW picture in aligned or weakly misaligned systems.
As the system evolves, the SF component grows, and its mass becomes comparable to that of the DW.
At late times, $M_{\rm SF}$ can exceed $M_{\rm DW}$, indicating that the outflow morphology transitions from a DW-dominated state to an SF-dominated state.

Figure~\ref{fig:tps-values}(b) shows the corresponding characteristic spatial extents of the two outflow components.
We plot the maximum vertical extent of the DW along the disk axis, $z_{\rm DW}$, and the maximum radial extent of the SF in the disk plane, $r_{\rm SF}$.
The DW initially opens a bipolar cavity along the $z'$-axis, and $z_{\rm DW}$ grows rapidly during the early phase.
Once a stable bipolar outflow has been established, the growth of $z_{\rm DW}$ becomes more gradual, and intermittent decreases in $z_{\rm DW}$ reflect phases in which the DW temporarily weakens or disappears in the misaligned system.
By contrast, $r_{\rm SF}$ increases steadily as the SF develops along the disk plane, although short episodes of decreasing $r_{\rm SF}$ occur when the expanding SF cavity is compressed by infalling envelope gas or when slower infalling material mixes into the SF component \citep[e.g.,][]{MachidaHosokawa2020}.
At late times, $r_{\rm SF}$ becomes comparable to or larger than $z_{\rm DW}$, demonstrating that the SF occupies a wide solid angle on spatial scales similar to or exceeding those of the main bipolar outflow.

Figures~\ref{fig:3dview_10panels} and \ref{fig:tps-values} show that in the misaligned model T60, a single protostellar system can drive two distinct outflow components from the same disk.
The system first launches a DW along the local rotation axis and subsequently develops a massive, extended SF roughly along the disk plane.
The relative contribution of the SF increases with time, and the system eventually enters a phase in which the SF dominates both the mass and the spatial extent of the outflow.

\begin{figure*}[t!]
  \begin{minipage}[t]{0.49\linewidth}
    \centering
    \includegraphics[width=\linewidth]{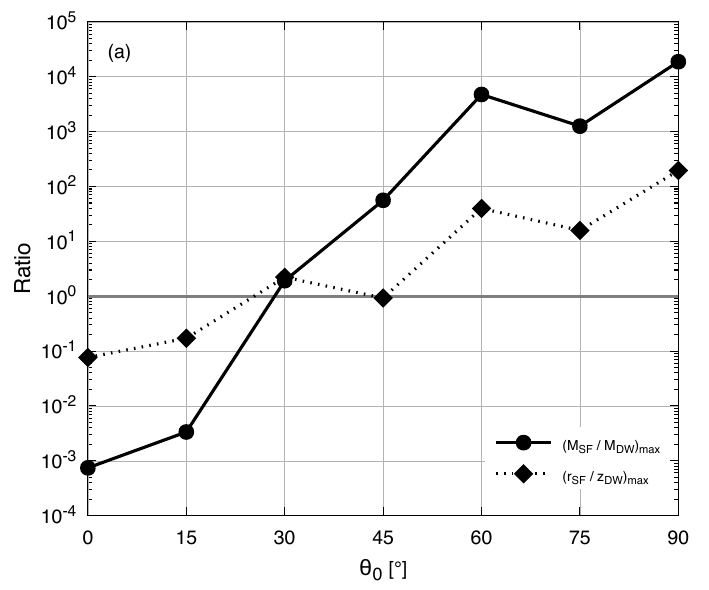}
  \end{minipage}
  \hfill
  \begin{minipage}[t]{0.49\linewidth}
    \centering
    \includegraphics[width=\linewidth]{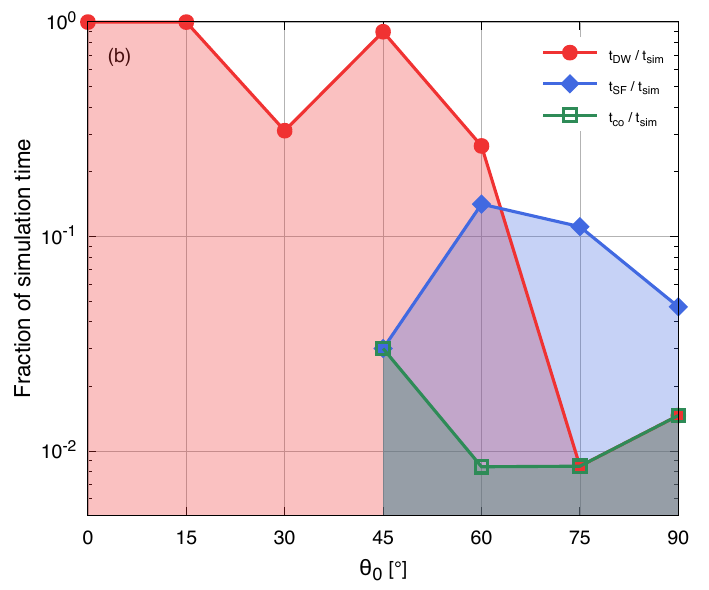}
  \end{minipage}
\caption{
Dependence of the relative importance of the SF component on the initial misalignment angle $\theta_0$.
Panel (a): Maximum mass ratio $(M_{\rm SF}/M_{\rm DW})_{\max}$ and maximum size ratio $(r_{\rm SF}/z_{\rm DW})_{\max}$ attained during the evolution for models T00--T90.
The horizontal gray line indicates unity.
Panel (b): Fractional durations of the simulation during which the DW mass, the SF mass, and both components simultaneously exceed a fiducial threshold of $10^{-3}\,\msun$.
The red, blue, and green filled regions indicate, respectively, the fractions of the simulated time for which DW, SF, and both components are ``on'', using the same mass threshold as in Figures~\ref{fig:tps-values}(b) and \ref{fig:tps-values_T00-90}(b).
}
\label{fig:theta-values}
\end{figure*}

\subsection{Dependence on the misalignment angle}
\label{sec:res:theta}

To quantify how the multiflow behavior depends on the initial misalignment, we analyze the full suite of models with $\theta_0$ ranging from $0^\circ$ to $90^\circ$.
For each model and each epoch, we compute the total mass in the two outflow components, $M_{\rm DW}(\tps)$ and $M_{\rm SF}(\tps)$, as well as characteristic spatial extents.
For the DW, we define the maximum vertical distance along the disk axis, $z_{\rm DW}$, while for the SF, we define the maximum radial extent in the disk plane, $r_{\rm SF}$.

We define an outflow component to be ``on'' when its instantaneous outflow mass exceeds a fiducial threshold, $M_{\rm out} > 10^{-3}\,\msun$.
Because our analysis focuses on the inner (near-source) outflow region with a characteristic extent of $L \sim 10^{3}$\,au (Figure~\ref{fig:tps-values}b), a relevant physical timescale is the flow crossing time, $t_{\rm flow} \sim L/v$.
Adopting typical outflow velocities for Class~0/I sources, $v \sim 10\,\kms$ \citep[e.g.,][]{Bally2016_ARAA,Tsukamoto2023PPVII}, gives $t_{\rm flow} \sim 500$\,yr for $L \sim 10^{3}$\,au.
Together with observed mass-loss rates of protostellar outflows, $\dot M_{\rm w}\sim10^{-6}\,\msunyr$ \citep[e.g.,][]{Podio2021,Dutta2024}, this implies a contemporaneous outflow-mass scale $\sim\!\dot{M}_{\rm w} t_{\rm flow} \sim 10^{-3}\,\msun$.
Thus, $10^{-3}\,\msun$ serves as a conservative operational criterion to identify sustained, dynamically significant activity and to suppress low-mass, short-lived fluctuations.

Figure~\ref{fig:tps-values_T00-90} extends the time-evolution analysis of Figure~\ref{fig:tps-values} to all misalignment angles.
Each row shows the temporal evolution of the stellar, disk, DW, and SF masses and the corresponding characteristic sizes for a given $\theta_0$, with the same colored background indicating phases during which the DW mass, the SF mass, and both components simultaneously exceed $10^{-3}\,\msun$.
This figure illustrates how the relative importance and spatial extent of the SF change systematically with $\theta_0$, but the detailed time-series information is cumbersome to compare across models.
Note that the simulations for $\theta_0 = 30^\circ$–$90^\circ$ were run up to $t_{\rm ps}\ge 10^5$ yr, while those for $\theta_0 = 0^\circ$ and $15^\circ$ were stopped at $t_{\rm ps} < 10^5$ yr due to extremely short timesteps encountered in the calculation.

To further quantify the relative importance of the circumstellar components in a way that is less sensitive to the absolute evolutionary times, Figure~\ref{fig:mps-mass_T00-90} plots the mass ratios of the disk, DW, and SF to the protostellar mass, as functions of protostellar mass for all models.
We find that the disk-to-protostar mass ratio remains nearly constant at $M_{\rm disk}/M_{\rm ps}\simeq 0.1$ throughout the evolution, showing no systematic dependence on the initial misalignment angle $\theta_0$.
In contrast, the DW contribution decreases with increasing $\theta_0$.
For nearly aligned systems ($\theta_0\le 30^\circ$), $M_{\rm DW}/M_{\rm ps}$ gradually declines with growth of the protostar and stays at the level of $\sim 0.1$.
For the intermediate case ($\theta_0=45^\circ$), $M_{\rm DW}/M_{\rm ps}$ is substantially reduced to $\sim 10^{-3}$--$10^{-2}$.
For strongly misaligned systems ($\theta_0\ge 60^\circ$), the DW mass fraction is suppressed over most of the evolution, remaining as low as $\sim 10^{-4}$.
The SF mass fraction is generally small, $M_{\rm SF}/M_{\rm ps}\ll 1$, in all models.
The epochs identified as strongly driven SF in Figure~\ref{fig:mps-mass_T00-90} correspond to transient enhancements of $M_{\rm SF}/M_{\rm ps}$, reaching $\gtrsim 10^{-3}$.

To provide a compact summary of these trends, Figure~\ref{fig:theta-values} condenses the information in Figure~\ref{fig:tps-values_T00-90} into a set of global diagnostics.
From the time evolution of $M_{\rm DW}$, $M_{\rm SF}$, $z_{\rm DW}$, and $r_{\rm SF}$ in each model (Figure~\ref{fig:tps-values_T00-90}), we extract the maximum mass ratio $(M_{\rm SF}/M_{\rm DW})_{\rm max}$ and the maximum size ratio $(r_{\rm SF}/z_{\rm DW})_{\rm max}$, as well as the fraction of the time during which each component is ``on'' according to the same $10^{-3}\,\msun$ threshold.
These diagnostics form the basis of the misalignment-angle trends discussed below.

Figure~\ref{fig:theta-values}(a) shows the maximum mass ratio $(M_{\rm SF}/M_{\rm DW})_{\rm max}$ and the maximum size ratio $(r_{\rm SF}/z_{\rm DW})_{\rm max}$ attained during the evolution.
For nearly aligned configurations with $\theta_0 < 30^\circ$, both ratios are much smaller than unity, indicating that the DW clearly dominates in both mass and spatial extent.
At intermediate misalignment, $\theta_0 \sim 30^\circ$, the SF becomes more prominent, and the mass and size ratios approach unity.
For large misalignment angles, $\theta_0 \gtrsim 45^\circ$, the SF can become more massive than the DW and reach a larger spatial extent, with both ratios exceeding unity.
These trends demonstrate the emergence of an SF-dominated outflow mode at a large misalignment angle.

Figure~\ref{fig:theta-values}(b) shows the fractional time during which each outflow component is strongly present, defined by a fiducial threshold of $10^{-3}\,\msun$.
For each model, we plot the fraction of the simulated time ($t_{\rm sim}$) during which the DW is on, the SF is on, and both components simultaneously exceed the threshold ($t_{\rm DW}$, $t_{\rm SF}$, and $t_{\rm co}$).
In aligned and weakly misaligned models, the DW remains active for most of the evolution, whereas the SF component is either absent or present only briefly.
As $\theta_0$ increases, the SF lifetime grows, and the system spends a substantial fraction of its evolution in an SF-dominated state for $\theta_0 \gtrsim 60^\circ$.
With the relatively high mass threshold adopted here, the period of strong co-existence of the two components is short.
However, tests with a lower threshold show that overlap phases become more frequent, albeit still brief.

The results show that this misalignment alters both the disk and the outflow \citep{Hirano2020}, and can generate additional outflow components beyond a single tilted axis.
A sufficiently large misalignment qualitatively changes the outflow morphology by enabling two distinct magnetically driven outflow channels from the same protostellar disk.
The classical DW dominates in nearly aligned systems, whereas an SF-dominated mode emerges at large misalignment and is consistent with the presence of secondary outflows observed in some protostellar systems.

\begin{figure}[t!]
\includegraphics[width=\linewidth]{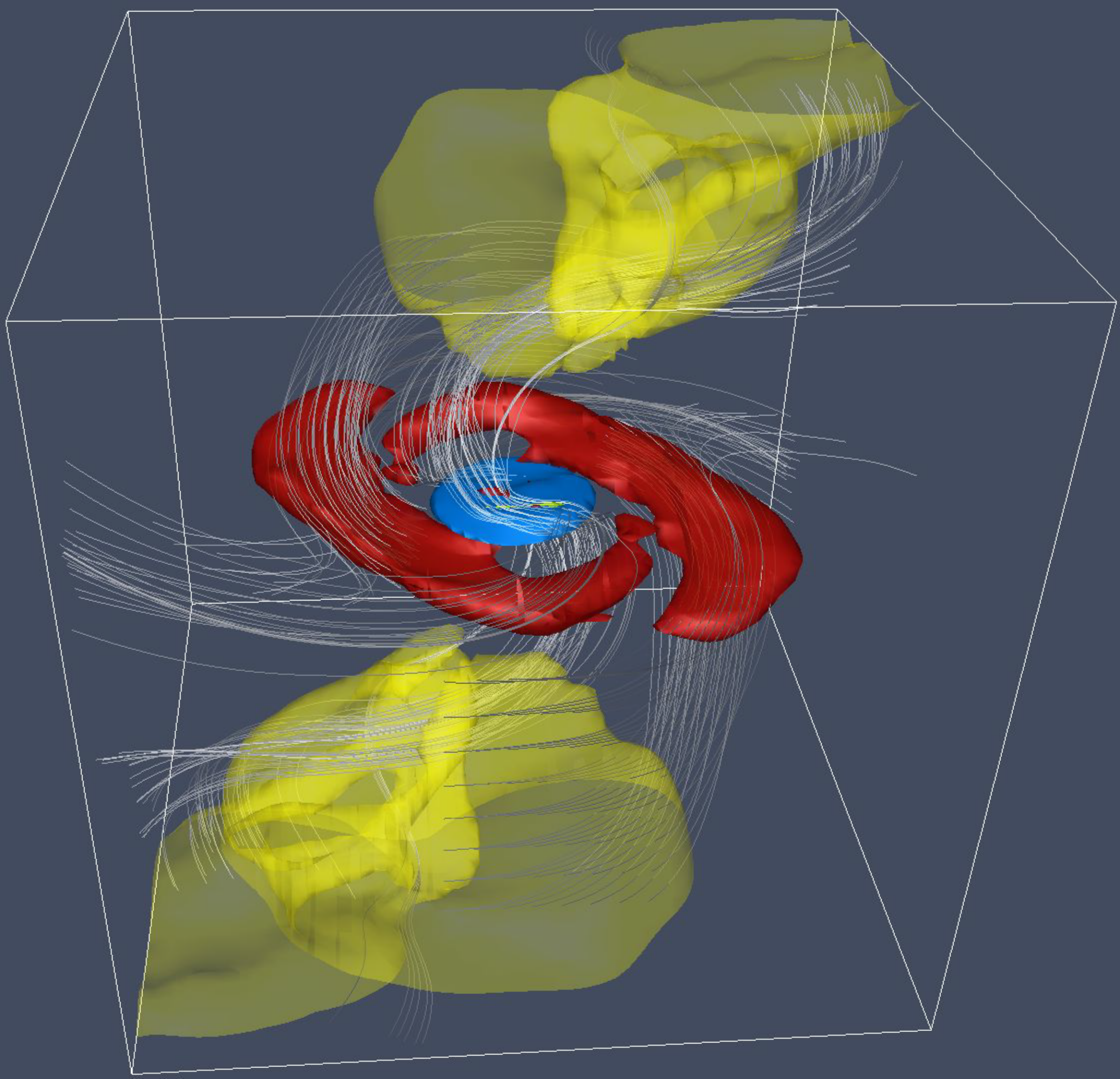}
\caption{
Simulation visualization of the misaligned protostellar system and the resulting multi-component outflows.  
A protostar is surrounded by a tilted circumstellar disk (blue).  
From the inner disk, a disk wind (DW; yellow lobes) is launched roughly along the local disk axis, following large-scale magnetic-field lines (gray curves).  
At the same time, twisted magnetic-field lines in and around the disk plane launch and accelerate a magnetically driven spiralflow (SF; red lobes) that propagates preferentially along the disk plane at a different position angle.
This configuration naturally produces two distinct outflow components from a single misaligned protostellar system, corresponding to the DW and SF components analyzed in this work.
}
\label{fig:mechanism}
\end{figure}

\section{Discussion} \label{sec:dis}

\subsection{Driving mechanism of SF outflow}
\label{sec:dis:mechanism}

Spiral-shaped, magnetically driven outflows have been reported in earlier simulations of misaligned, magnetized cores. 
\citet{MatsumotoHanawa2011} identified a distinct SF outflow in collapsing turbulent cores and showed that it is associated with a twisted magnetic-field configuration in which the rotation axis is inclined with respect to the large-scale field. 
\citet{Machida2006} found a similar internal field morphology for inclined rotators.
However, previous simulations were not long enough to study the spiral outflow's long-term behavior. 
\citet{Matsumoto2017} further argued that internally twisted field lines in misaligned systems can cause the outflow to deviate from a simple bipolar shape and contribute to a spiral-shaped flow component.

Our simulations confirm this basic picture and place it within a unified framework alongside the DW. 
Figure~\ref{fig:mechanism} schematically illustrates how a large initial misalignment between the core angular momentum and the magnetic field leads to the magnetic configuration that drives the SF component.
In our misaligned models, the large-scale magnetic field is initially inclined with respect to the core angular-momentum vector. 
As the collapse proceeds and a circumstellar disk forms, the inner footpoints of the magnetic field lines become anchored in the rotating disk and are twisted by disk rotation.
The outer footpoints are anchored near the disk's outer edge and remain coupled to the ambient infalling gas, which also shears and twists the field.
This differential twisting produces two families of field lines with distinct geometries: some bend away from the disk and open toward the polar regions, whereas others remain strongly wrapped and bent within or near the disk plane.

Gas loaded onto the former family is accelerated magnetocentrifugally along the inclined field lines, forming the classical DW.
Gas attached to the latter family, in which the magnetic field lines have a spiral configuration, is likewise accelerated by rotation, forming an SF.
In many snapshots, the SF appears as an approximately point-symmetric, two-armed pattern in the disk plane.
This can be understood if the SF is launched episodically from the most strongly wound, disk-plane--parallel field lines: as the twisting builds up, only a limited number of azimuthal sectors dominate the launching.
With finer time sampling, additional weaker arms can appear, but the morphology is typically dominated by two main arms.
For the latter family, the configuration of the magnetic field lines and the associated outflow is qualitatively similar to the Parker spiral considered in the context of the solar wind \citep{Parker1958,Weber1967,Spruit1996}.
However, in the case of the solar wind, the thermal pressure gradient force plays a more important role in accelerating the gas than the Lorentz force.
Thus, although the magnetic-field and outflow patterns are similar between the Parker spiral and those found in our simulations and in previous studies \citep[e.g.,][]{Matsumoto2017}, the underlying driving mechanisms differ.
In summary, a strongly misaligned system provides two distinct magnetic ``flows'' within a single protostellar disk: one that flings material along the rotation axis (DW), and another that transports mass and angular momentum along the disk plane (SF).

A key new result of this work is that we follow both components throughout protostellar evolution and quantify how they coexist and compete over time. 
Whereas previous studies demonstrated that SFs can emerge in misaligned collapse \citep[e.g.,][]{MatsumotoHanawa2011,Matsumoto2017}, our analysis shows that the relative importance and lifetime of the SF are controlled by the misalignment angle $\theta_0$, and that for large $\theta_0$ the SF can become more massive and more extended than the DW. 
The present simulations include only Ohmic dissipation, whereas ambipolar diffusion is expected to be more important in the lower-density region.\footnote{A practical reason for adopting the Ohmic-only setup in this study is that including ambipolar diffusion, as well as a more exact treatment of the Ohmic term, requires substantially smaller timesteps for numerical stability, which makes it difficult to follow the long-term evolution and durations of the DW and SF within a feasible computational cost.}
Therefore, although our results show that a misaligned system can develop a prominent SF component, the quantitative impact of ambipolar diffusion on the SF strength, lifetime, and launching radius remains uncertain and should be examined in future three-dimensional non-ideal MHD simulations that include ambipolar diffusion.

Observational evidence indicates that large misalignment angles between magnetic fields and outflows or disks are common rather than exceptional.
Dust-polarization surveys on core scales have shown broad or even bimodal distributions of the angle between the magnetic field and the outflow, consistent with preferential misalignment or nearly random orientations rather than tight alignment \citep[e.g.,][]{Hull2013, Chapman2013}.
Nonideal MHD simulations of core and filament evolution found that the distribution of the angle between the angular-momentum vector and the magnetic field remains broad down to the onset of collapse \citep[e.g.,][]{Machida2020, Misugi2024}.
These results imply that the large misalignment angles required for the SF–dominated mode identified here should be realized in a non-negligible fraction of star-forming cores.

\subsection{Time-dependent stress diagnostics of the DW and SF in model T60}
\label{sec:dis:stress}

\begin{figure*}[t!]
\centering
\includegraphics[width=\linewidth]{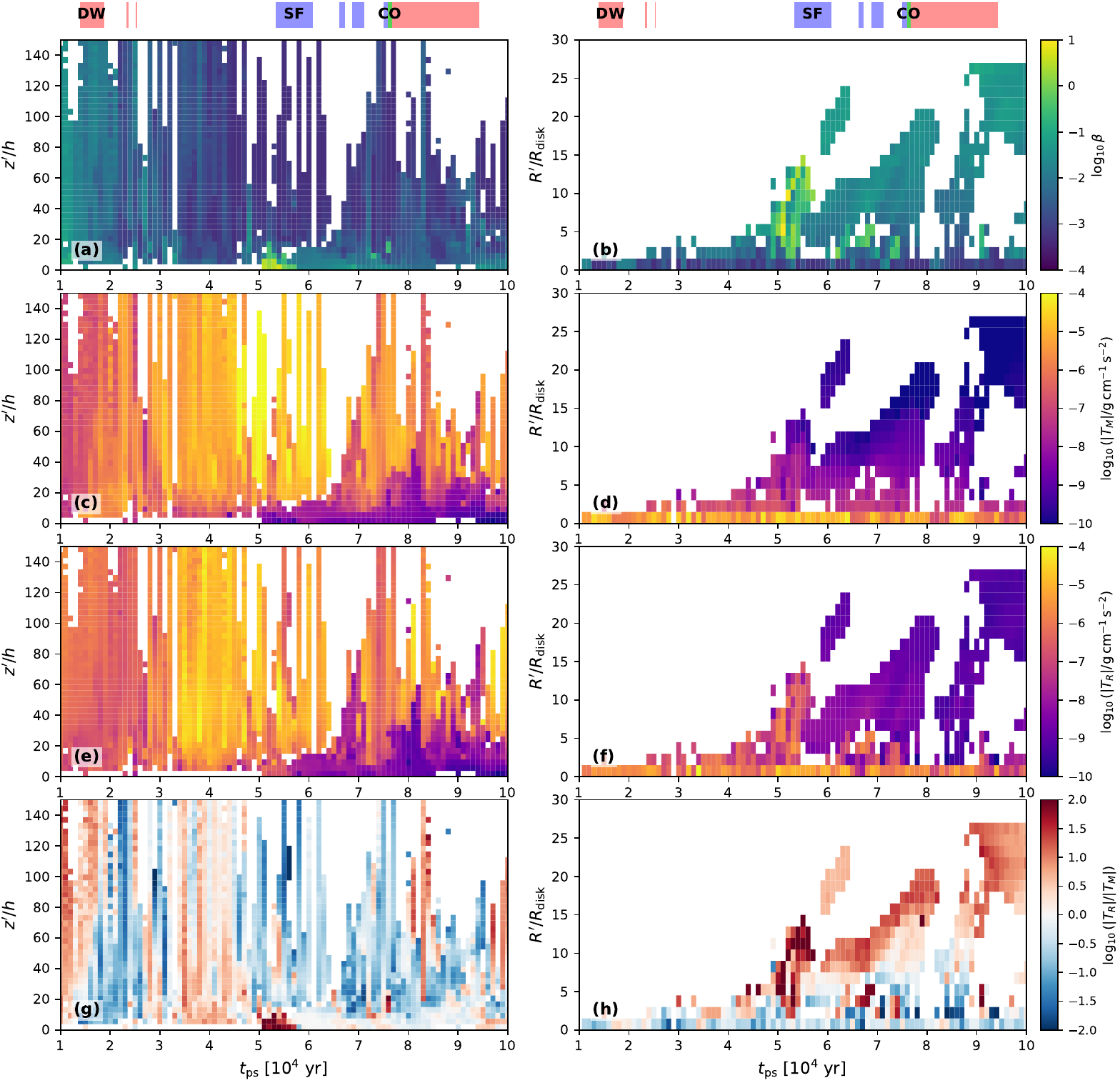}
\caption{
Disk-frame diagrams showing the time evolution of the DW (left) and SF (right) in the representative model T60.
Panels (a) and (b) show the plasma beta, panels (c) and (d) the Maxwell stress, panels (e) and (f) the Reynolds stress, and panels (g) and (h) the stress ratio $\log_{10}(|T_R|/|T_M|)$.
The left column is plotted against $z'/h$ for the DW, and the right column against $R'/R_{\rm disk}$ for the SF.
White regions indicate bins with too few cells satisfying the DW or SF selection criteria to be shown reliably.
The colored strips above the top panels indicate the intervals in which the DW, SF, or both components are active.
}
\label{fig:stress-timeseries}
\end{figure*}

To examine whether the DW and SF correspond to distinct dynamical regimes, we additionally analyze the time evolution of the plasma beta and of the Maxwell and Reynolds stresses in the disk frame.
Here, all quantities are evaluated in the instantaneous disk frame whose $z'$-axis is aligned with the angular-momentum vector of the dense disk; $R'=(x'^2+y'^2)^{1/2}$ and $z'$ are the cylindrical coordinates in this frame, $R_{\rm disk}$ is the maximum $R'$ of cells satisfying our disk criteria, $h=c_s/\Omega_K(R')$ is the local disk scale height, $B_{z'}$ and $B_{\phi'}$ are the vertical and azimuthal magnetic-field components, and $\delta v_{z'}$ and $\delta v_{\phi'}$ are residual velocities obtained after subtracting the density-weighted mean $v_{z'}(R')$ and $v_{\phi'}(R')$ profiles at each snapshot.
At each snapshot, the plotted value in each $z'/h$ or $R'/R_{\rm disk}$ bin is taken as the median over all cells satisfying the DW or SF selection criteria in that bin.
The plasma beta is defined as
\begin{equation}
\beta = \frac{8\pi \rho c_s^2}{B^2}.
\end{equation}
For the DW we use the vertical transport components
\begin{equation}
T_M^{z\phi} = -\frac{B_{z'}B_{\phi'}}{4\pi}, \qquad
T_R^{z\phi} = \rho\,\delta v_{z'}\,\delta v_{\phi'}.
\end{equation}
For the SF we use the radial transport components
\begin{equation}
T_M^{R\phi} = -\frac{B_{R'}B_{\phi'}}{4\pi}, \qquad
T_R^{R\phi} = \rho\,v_{R'}\,\delta v_{\phi'}.
\end{equation}
In addition to the absolute values of the Maxwell and Reynolds stresses, we plot their ratio in order to identify which stress dominates the angular-momentum transport in each outflow component.
Although this stress decomposition does not provide a unique force decomposition of the outflow acceleration, it serves as a useful diagnostic of whether the DW and SF are associated with different transport regimes.
Figure~\ref{fig:stress-timeseries} presents the resulting disk-frame diagrams for the representative T60 model.

Panels (a) and (b) show that both outflow components satisfy $\beta < 1$ over most of the active regions, confirming that the dynamics are generally magnetically dominated.
In addition, the stress maps reveal that the two components are dynamically distinct.
In the DW, the Maxwell and Reynolds stresses shown in panels (c) and (e) vary strongly with time, and panel (g) exhibits alternating intervals of Maxwell-dominated and Reynolds-dominated transport.
This indicates that the DW in the misaligned T60 model is not described by a single stationary stress balance, but instead evolves between different transport states during the protostellar phase.

The SF shows a different stress signature.
Panels (d) and (f) indicate that the SF maintains a strong stress signal over an extended range in $R'/R_{\rm disk}$.
In particular, panel (h) shows that the launching region is characterized by Maxwell-stress-dominated transport, consistent with the winding-up of magnetic field lines near the base of the flow, while over a broader range of $R'/R_{\rm disk}$ the Reynolds stress also makes a substantial contribution during the main SF-active epochs.
Thus, the DW and SF are associated with systematically different stress regimes.
This difference supports the interpretation that the SF is not merely a geometrically broadened extension of the classical DW, but a dynamically distinct outflow component produced by the twisted magnetic structure associated with the disk--field misalignment.
We note that a self-gravitational stress tensor could not be reconstructed from the available outputs, and therefore Figure~\ref{fig:stress-timeseries} focuses on Maxwell and Reynolds stresses together with $\beta$.

\subsection{Implications for observed multiple outflows}
\label{sec:dis:obs}

High-angular-resolution ALMA observations have revealed multiple outflow components in several low-mass protostars, where a secondary flow-like structure appears at a different position angle and with distinct kinematics from the main bipolar outflow \citep[e.g.,][]{Okoda2021,Sato2023,Sai2024}.  
These secondary components are not always easily explained by multiple protostars, a single precessing jet, or a purely wide-angle wind, suggesting that more complex launching mechanisms may be at work.

Our simulations provide a concrete physical framework in which multiple outflow components naturally arise from a single protostellar system.  
In misaligned models with large initial angles $\theta_0$, the system simultaneously drives a DW roughly along the disk rotation axis and a SF component launched roughly along the disk plane.
As shown in Figures~\ref{fig:tps-values} and \ref{fig:theta-values}, the DW and SF are highly time-variable and can appear episodically: the system alternates between DW-dominated and SF-dominated phases, with intervals during which both components coexist.
For appropriate lines of sight, these components would be projected onto the sky plane at different position angles and characteristic velocities, resembling the observed combination of a main outflow and a secondary misaligned outflow or a fossil remnant of such an episode.

Our models also make several qualitative predictions that can be tested with future observations.
First, the SF component is expected to be more extended in the disk plane than in the polar direction, and to be associated with a twisted magnetic-field geometry, possibly detectable via polarization or Zeeman observations.
Second, the relative prominence of the secondary flow should correlate with signatures of disk or envelope misalignment, as well as with indications of strong magnetic twisting on scales of a few hundred au.
Observationally, disk-outflow misalignment appears to be common: \citet{Feeney-Johansson2026} find that 7 of 19 eDisk Class~0/I sources show outflow axes that are not perpendicular to their disks (magnetic-field orientations are not constrained).
Such systems also exhibit complex disk morphologies, including warps and strongly tilted disk planes, indicating that large misalignment angles are present in real sources \citep[e.g.,][]{Maury2022review}.
Finally, because the SF becomes increasingly crucial at relatively late times in the evolution of the T60 model, secondary flows may be more common in systems where a substantial circumstellar disk has already formed.
If such trends are confirmed, they would support a unified picture in which misaligned magnetic fields and disks naturally produce multiflow outflows, linking the SFs seen in simulations to the multiple outflows or their fossil remnants observed with ALMA.

\begin{acknowledgments}
This work used the computational resources of the HPCI system provided by the supercomputer system SX-Aurora TSUBASA at Tohoku University Cyberscience Center and Osaka University Cybermedia Center through the HPCI System Research Project (Project ID: hp220003, hp230035, and hp240010) and Earth Simulator at JAMSTEC provided by 2022, 2023, and 2024 Koubo Kadai.
Numerical analyses were carried out on analysis servers at the Center for Computational Astrophysics, National Astronomical Observatory of Japan.
This work was supported by JSPS KAKENHI Grant Numbers JP21K13960 and JP21H01123 / JP23K20864 (S.H.), JP18H05222, JP20H05847, and JP24K00674 (Y.A.), and JP25K07369 (M.N.M).
This work was also supported by a NAOJ ALMA Scientific Research grant (No. 2022-22B).
\end{acknowledgments}

\begin{contribution}
All authors contributed equally to the manuscript.
\end{contribution}

\software{Astropy \citep{astropy1_2013, astropy2_2018, astropy3_2022}, 
          h5py \citep{h5py_2023},
          Matplotlib \citep{matplotlib2007},
          NumPy \citep{numpy2020}
          }

\appendix
\restartappendixnumbering

\section{Order-of-magnitude estimate of the Ohmic approximation error}
\label{sec:app:ohmic}

As noted in Section~\ref{sec:method_sim}, the Ohmic term is implemented in the form $\eta_{\rm O}\nabla^2\mathbf{B}$.
For spatially varying $\eta_{\rm O}$ and under the condition $\nabla\cdot\mathbf{B}=0$, the full operator can be written as
\begin{equation}
-\nabla\times(\eta_{\rm O}\nabla\times\mathbf{B})=\eta_{\rm O}\nabla^2\mathbf{B}-\nabla\eta_{\rm O}\times(\nabla\times\mathbf{B}).
\end{equation}
We therefore estimate the magnitude of the neglected term using a one-dimensional post-processing analysis along representative cuts through the outflowing gas.

We define
\begin{eqnarray}
  T_{\rm lap}(s) = \left|\eta_{\rm O}\frac{d^2B}{ds^2}\right| , \\
  T_{\rm grad}(s) = \left|\frac{d\eta_{\rm O}}{ds}\frac{dB}{ds}\right| , \\
  R(s) = T_{\rm grad}(s) / T_{\rm lap}(s) .
\end{eqnarray}
where $T_{\rm lap}(s)$ represents the retained Laplacian term, $T_{\rm grad}(s)$ the neglected gradient term, and $R(s)$ the relative magnitude of the approximation error.
Here, $s$ denotes the coordinate along the one-dimensional cut through the outflowing gas, with $s=z'$ for the DW and $s=R'$ for the SF.

\begin{figure}[t!]
  \centering
  \includegraphics[width=0.95\linewidth]{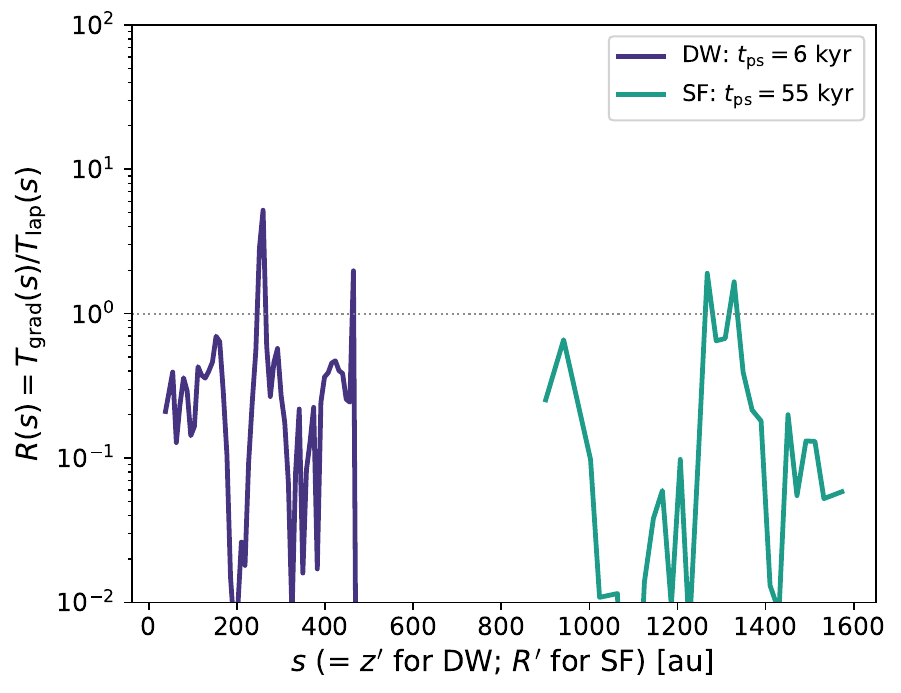}
  \caption{
  Order-of-magnitude estimate of the error introduced by approximating the Ohmic term as $\eta_{\rm O}\nabla^2\mathbf{B}$.
  The plotted quantity is $R(s)=T_{\rm grad}(s)/T_{\rm lap}(s)$, where $T_{\rm lap}(s)=|\eta_{\rm O}d^2B/ds^2|$ is the retained term and $T_{\rm grad}(s)=|(d\eta_{\rm O}/ds)(dB/ds)|$ is the neglected term.
  The figure compares a representative DW profile at $\tps=6$\,kyr and the outer-tail SF profile at $\tps=55$\,kyr in the T60 model, with $s=z'$ for the DW and $s=R'$ for the SF.
  }
  \label{fig:app:ohmic_error}
\end{figure}

Figure~\ref{fig:app:ohmic_error} shows representative profiles in the T60 model.
For the DW, we use the upper-hemisphere outflow cells at $\tps=6$\,kyr and evaluate the profile as a function of $z'$.
For the SF, we use the outer-tail SF cells at $\tps=55$\,kyr and evaluate the profile as a function of $R'$.
The Ohmic resistivity is reconstructed using the same prescription as in the simulation, and the profiles are binned and smoothed before numerical derivatives are taken.

For both the representative DW and SF cuts shown here, $R(s)$ locally exceeds unity in a limited part of the profile, but remains generally of order unity or smaller over most of the displayed range.
This indicates that in the representative outflow regions examined here, the neglected gradient term is not systematically larger than the retained Laplacian term, suggesting that the use of the approximate Ohmic operator is reasonable for the present analysis.

\bibliography{ms}{}
\bibliographystyle{aasjournalv7}





\end{document}